\documentclass[runningheads]{llncs}
\usepackage[T1]{fontenc}
\usepackage{xcolor}

\usepackage{graphicx}
\usepackage{soul}
\usepackage{booktabs,siunitx}
\usepackage{multirow}
\usepackage{array}
\usepackage{caption}
\usepackage{booktabs}
\usepackage{geometry}
\usepackage{subfigure}
\usepackage{amsmath}
\usepackage{amssymb}

\usepackage{color}
\usepackage{hyperref}
\begin{document}
\title{VIMs: Virtual Immunohistochemistry Multiplex staining via Text-to-Stain Diffusion Trained on Uniplex Stains}
\titlerunning{VIMs: Virtual Immunohistochemistry Multiplex staining}
\authorrunning{S. Dubey et al.}
 \vspace{-0.1in}
\author{Shikha Dubey\inst{1,2}, Yosep Chong\inst{3,4} \and Beatrice Knudsen\inst{3,5} \and
Shireen Y. Elhabian \inst{1,2,6} }
\institute{Kahlert School of Computing, University of Utah, USA \and
Scientific Computing and Imaging Institute, University of Utah, USA \and
The Catholic University of Korea College of Medicine, S. Korea \and
Huntsman Cancer Institute, University of Utah Health, USA \and
Department of Pathology, University of Utah, USA \and Corresponding author\\ \{\{{shikha.d,shireen\}@sci, beatrice.knudsen@path\}.utah.edu}, ychong@catholic.ac.kr}
\maketitle 
\vspace{-0.18in}
\begin{abstract}
This paper introduces a Virtual Immunohistochemistry Multiplex staining (VIMs) model designed to generate multiple immunohistochemistry (IHC) stains from a single hematoxylin and eosin (H\&E) stained tissue section. 
IHC stains are crucial in pathology practice for resolving complex diagnostic questions and guiding patient treatment decisions. While commercial laboratories offer a wide array of up to 400 different antibody-based IHC stains, small biopsies often lack sufficient tissue for multiple stains while preserving material for subsequent molecular testing. This highlights the need for virtual IHC staining. Notably, VIMs is the first model to address this need, leveraging a large vision-language single-step diffusion model for virtual IHC multiplexing through text prompts for each IHC marker. VIMs is trained on uniplex paired H\&E and IHC images, employing an adversarial training module. Testing of VIMs includes both paired and unpaired image sets. To enhance computational efficiency, VIMs utilizes a pre-trained large latent diffusion model fine-tuned with small, trainable weights through the Low-Rank Adapter (LoRA) approach. Experiments on nuclear and cytoplasmic IHC markers demonstrate that VIMs outperforms the base diffusion model and achieves performance comparable to Pix2Pix, a standard generative model for paired image translation. Multiple evaluation methods, including assessments by two pathologists, are used to determine the performance of VIMs. Additionally, experiments with different prompts highlight the impact of text conditioning. This paper represents the first attempt to accelerate histopathology research by demonstrating the generation of multiple IHC stains from a single H\&E input using a single model trained solely on uniplex data. This approach relaxes the traditional need for multiplex training sets, significantly broadening the applicability and accessibility of virtual IHC staining techniques.
\vspace{-0.1in}
\end{abstract}
\vspace{-0.34in}
\section{Introduction} \label{sec:intro}
\vspace{-0.1in}
Pathologists begin their microscopic assessments of tissues using hematoxylin and eosin (H\&E) stained tissue sections. However, challenging cases require more precise information that cannot be delivered solely by H\&E staining. This additional information is provided through the staining of tissues with antibodies using a method called immunohistochemistry (IHC). Antibodies label specific proteins in the nucleus, cytoplasm, or membrane of cells, and their binding is visualized by a brown color in the tissue. The antibody stain is easily distinguished from the blue Hematoxylin stain, which highlights DNA and RNA to reveal cellular organization. IHC staining, though informative, is time-consuming and expensive, often taking hours or days. Standard IHC methods rely on staining with one antibody at a time (uniplex stain). However, multiple stains are needed to arrive at the correct diagnosis, requiring multiple tissue sections \cite{ihcintor}. Small biopsies contain insufficient tissue for multiple IHC assays and cannot provide sufficient material for proper diagnostic workup and subsequent molecular analysis for treatment selection. Standard multiplexed IHC, which consists of staining the same tissue section with multiple antibodies, would solve the problem of limited tissue availability. However, slides stained by multiplexed IHC cannot be easily analyzed by regular light microscopy. In addition, the number of IHC assays that can be performed in the same tissue section is limited by overlapping staining patterns of antibodies. Each antibody is labeled with a unique color, necessitating color deconvolution of multiplexed IHC images. While a seemingly simple task, algorithmic color unmixing of IHC images constitutes a problem \cite{knudsen}. Altogether, generating multiple virtual IHC stains from an H\&E image would provide a benefit to patients who have only limited tissue samples for diagnostic workup and treatment planning or who live in countries where IHC resources are not available.  

Advancements in vision-based image generation using generative adversarial network (GAN) models, such as Pix2Pix \cite{pix2pix} and CycleGAN \cite{cyclegan}, have inspired their application in the medical domain, specifically in histopathology \cite{rivensonphasestain,autoflorrivenson,vivo,zingman2023comparative,stainnet,staingan,bci,scgan,cycleganvariants,pix2pixvar2,restaining}. These models have been employed for tasks such as virtual H\&E and Masson’s Trichrome staining  \cite{zingman2023comparative} and converting autofluorescence images to virtual HER2 \cite{autoflorrivenson}, stain normalization \cite{stainnet,staingan}, and virtual IHC generation from H\&E images \cite{bci,scgan,cycleganvariants,pix2pixvar2,restaining}. So far, virtual IHC staining models, trained on paired or unpaired data, typically generate a uniplex or single IHC stain from each H\&E image. Paired IHC and H\&E data are difficult to generate due to the need for pixel-wise registration of whole-slide images (WSIs) H\&E to IHC, whereas unpaired data do not require such registration. Generating multiple IHC markers from a single H\&E sample remains relatively unexplored and challenging, though a few recent GAN-based models \cite{unpairedmultidomain,multidomain2,MultiVStain} have attempted to address this issue. While models in \cite{unpairedmultidomain,multidomain2} are trained independently for different IHC markers without learning proper associations among IHCs, Multi-VSTAIN \cite{MultiVStain} requires multiplexed paired data for training, which are difficult to obtain. Therefore, GAN-based models reduces their practical application. The proposed model, Virtual Immunohistochemistry Multiplex staining (VIMs), does not need multiplex paired training data and trains the model in an end-to-end manner, taking advantage of associations among IHCs.

Recently, conditional diffusion models (DMs) \cite{latentsapcemodel,Controlnet,im2im-Turbo,Im2ImDiffuse,SD-Turbo,text2image}, a class of generative models, have demonstrated remarkable success in image synthesis tasks. These models operate by learning to reverse a gradual noising process, enabling the generation of high-quality images. However, traditional DMs \cite{latentsapcemodel,Controlnet,Im2ImDiffuse} require multiple denoising steps during inference, which can be time-consuming. Despite this, DMs produce higher fidelity images compared to conventional GAN-based models. This is crucial in the medical domain, where detailed cellular-level morphological information is vital for accurate disease diagnosis. Large vision-language diffusion models (LVDMs) are generally challenging to train from scratch, and fine-tuning their billions of parameters requires extensive computational power and large datasets. Recent advancements, such as adapters \cite{LoRA} and fine-tuning methods \cite{Controlnet}, have improved their training efficiency. Additionally, optimized DMs \cite{im2im-Turbo,SD-Turbo} developed to reduce the number of steps needed during inference. However, the potential of these optimized DMs for virtual IHC staining has not yet been explored. This paper aims to fill this gap by investigating the use of such advanced DMs for multiplex IHC marker generation. Inspired by recent advances incorporating adversarial objectives into conditional DMs \cite{SD-Turbo,im2im-Turbo}, this paper introduces VIMs. VIMs leverages a text-conditioned single-step DM to generate multiple paired virtual IHC images from the same H\&E image, allowing multiple IHC markers to be visualized in the exact same cells, which is difficult to obtain from slides used in clinical practice. Additionally, evaluating generative models, especially in medical imaging, is challenging \cite{scgan}; thus, this study uses multiple evaluation methods, including a manual assessment by two pathologists.

\vspace{0.05in}
\noindent The primary contributions of this study are as follows:
\vspace{-0.07in}
\begin{enumerate}
    \item Introduces VIMs 
    \footnote{We will release all the code related to the paper at a future date for public usage.}, a method for virtual multiplexed IHC that generates multiple IHC stains from a single H\&E image using a text-conditioned single-step DM. VIMs is the first to adapt such a model for virtual IHC staining, incorporating adversarial learning objectives to preserve tissue structures during translation.
    \item VIMs is trained on uniplex paired (pixel-level registered H\&E to IHC) data with pathologist-validated prompts, using a large latent DM with small trainable weights via the Low-Rank Adapter (LoRA) for efficient end-to-end training.
    \item Extensive experiments are conducted on two gland markers, CDX2 and CK8/18, using both paired and unpaired test sets. Two pathologists confirmed the high performance of VIMs in both settings. Different prompts were tested to highlight the impact of text conditioning. For the first time, DICE and mIoU metrics were used to evaluate whether VIMs accurately stains the correct cells and places the staining correctly in the nucleus or cytoplasm.
\end{enumerate}
\vspace{-0.10in}
The paper covers VIMs methodology in Section \ref{sec:method}, experimental and results in Section \ref{sec:exp}, and conclusions and future directions in Section \ref{sec:conclusion}.
\begin{figure}[t]
    \centering
    \includegraphics[clip, trim=0.2cm 2.70cm 0.1cm 1.8cm, scale=0.39]{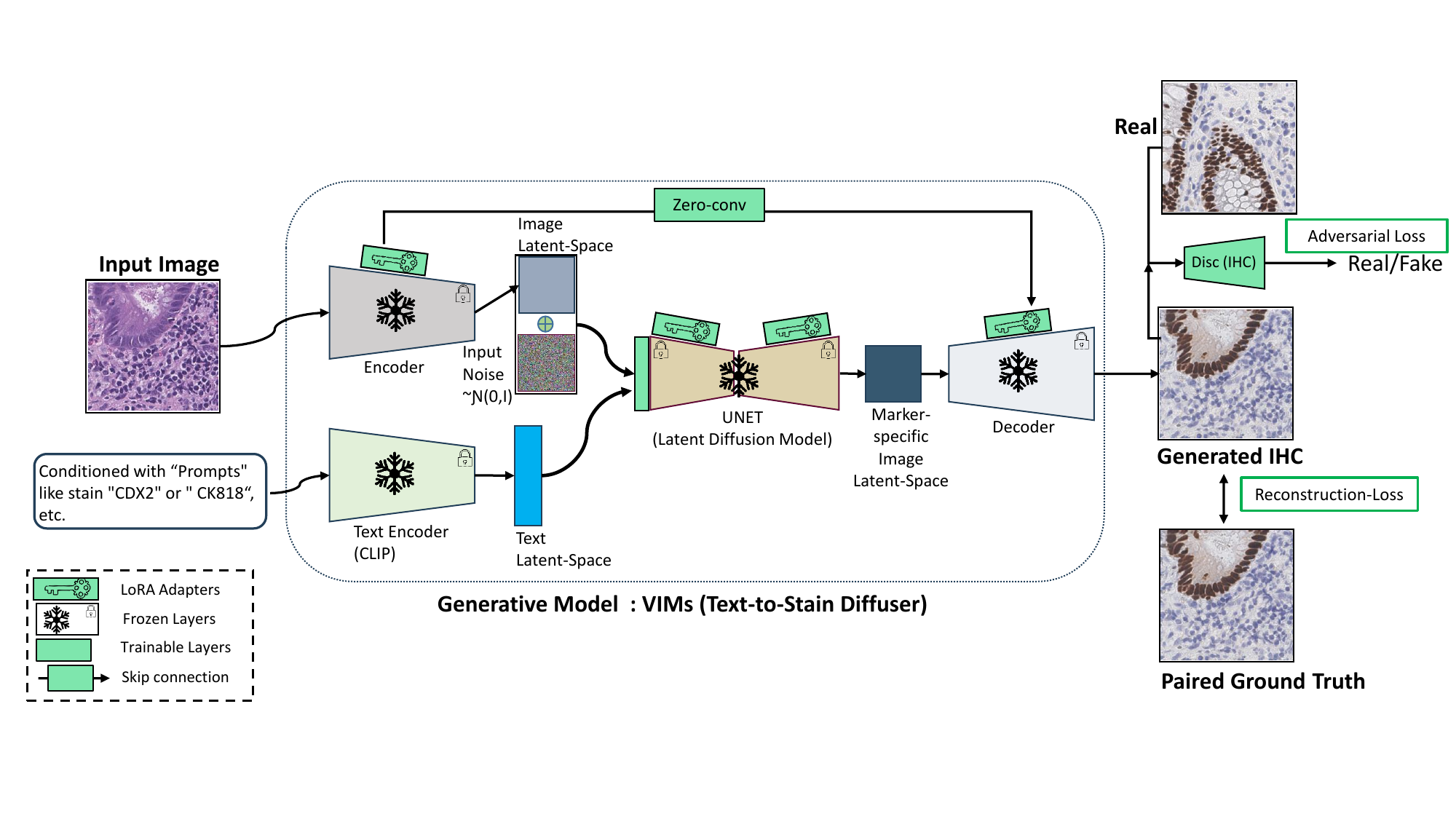}  
    \vspace{-0.05in}
    \caption{\textbf{VIMs: Proposed Multiplex IHC Staining Model.} The pre-trained LDM, one of the Large Language Models (LLMs), is optimized for the virtual multiplex IHC generation task with a minimal number of trainable parameters.}
    \label{fig:model}
    \vspace{-0.12in}
\end{figure}
 \vspace{-0.12in}

\section{Methods: Virtual Immunohistochemistry Multiplex staining (VIMs)} \label{sec:method} %\label{subsec:vims}
\vspace{-0.1in}
The proposed model, VIMs (Fig. \ref{fig:model}), is adapted from recent advancements in conditional DMs \cite{SD-Turbo,Controlnet,im2im-Turbo,latentsapcemodel} and applies an LVDM to the IHC image generation task. VIMs consists of multiple pre-trained and trainable modules: a pre-trained text encoder, Contrastive Language-Image Pretraining (CLIP) \cite{clip} to obtain the text latent space; an encoder to obtain the input H\&E image latent space; a denoising UNet block \cite{latentsapcemodel}; and a decoder to generate the IHC stain image from the marker-specific image latent space (the output of the denoising UNet block). Gaussian noise is added before passing the features from the latent space of the input H\&E image through the denoising UNet block of the latent diffusion model (LDM). This UNet also receives input from the encoded prompt/text. The adapter LoRA \cite{LoRA} and techniques from \cite{Controlnet} are incorporated into the framework of the model (refer to Fig.\ref{fig:model}) for efficient training. Additionally, adversarial learning \cite{pix2pix}, combined with reconstruction loss, ensures high-fidelity image reconstruction and accurate domain translation. The details of each module and the training of VIMs are provided in the following subsections.
\vspace{-0.09in}
\subsection{Latent Diffusion Model (LDM): Encoder-UNet-Decoder} \label{subsec:ldm}
\vspace{-0.05in}
VIMs utilizes a pre-trained image encoder, denoising UNet, and decoder from LDM \cite{latentsapcemodel}. Unlike \cite{Controlnet} (comparison shown in Figs. \ref{fig:qual:ck818} and \ref{fig:qual:cdx2}, Table \ref{tab:comparison_paired_models}), the embeddings of the input image are directly combined with the noise maps, rather than being conditioned through the decoder of the denoising UNet. This approach avoids conflicts between the noise map and the input image domain, preserving input details for generating associated IHC stains \cite{im2im-Turbo}. Most layers of the LDM-UNet are frozen, except for the first layer, which remains trainable for feature preservation during fine-tuning. VIMs uses the pre-trained encoder ($E$) to encode the H\&E input image into the latent space ($x = E(H)$) and the decoder ($D$) to decode the denoised marker-specific latent space to generate the IHC image ($I = D(y)$). The diffusion process progressively denoises the combination of the input H\&E image latent space and a noise variable ($z_t$), sampled from a Gaussian distribution $\mathcal{N}(0, I)$. Additionally, the denoising is conditioned on text prompts until a clean, denoised IHC image latent space is generated. This process is represented as:
\vspace{-0.1in}
\begin{equation}
    y_i = Q(x_i + z_t, c_i)
    \label{eq1}
\end{equation}
\vspace{-0.02in}
where $Q$ maps the $i_{th}$ input domain H\&E sample's latent space ($x_i$), combined with the noise map ($z_t$), to the associated specific IHC latent space ($y_i$) based on the input prompt ($c_i$).
\vspace{-0.09in}
\subsection{Text/Prompt Encoder}\label{subsec:prompt}
\vspace{-0.05in}
The text/prompt encoder plays a vital role in VIMs, enabling the generation of multiple IHC images that represent different antibody IHC stains from a single H\&E image through text conditioning. VIMs maps the H\&E latent space to marker-specific IHC latent spaces and learns correlations among different markers based on the input prompt. As an end-to-end model, VIMs is trained on a uniplex H\&E and IHC dataset using the prompt encoder, thereby eliminating the need for multiplex paired data (input images stained and registered at the pixel level with multiplex IHC). We evaluated the impact of different types of prompts (refer Suppl. Fig.\ref{fig:prompts}) on VIMs performance (refer Suppl. Table \ref{tab:comparison_paired_prompt}, Fig.\ref{fig:qual:cdx2}) during training. Additionally, VIMs was compared with a base model \cite{latentsapcemodel}, where IHC generation is conditioned only on the input H\&E image with an empty prompt to assess the general effect of text conditioning (refer Tables \ref{tab:comparison_paired_models} and \ref{tab:comparison_unpaired}, Figs.\ref{fig:qual:ck818}, \ref{fig:qual:cdx2}). VIMs utilizes the pre-trained CLIP model \cite{clip} as the text/prompt encoder, which encodes textual descriptions $p$ into feature vectors $t_p$ that guide the image generation process. This enables VIMs to generate specific types of IHC stains based on user-provided text prompts, ensuring that the output aligns with the desired staining characteristics.
\vspace{-0.09in}
\subsection{LoRA Adapter and Skip Connections}\label{subsec:lora}
\vspace{-0.05in}
To efficiently adapt the pre-trained LDM \cite{latentsapcemodel,im2im-Turbo,SD-Turbo} for the virtual IHC staining task and integrate new functionalities without overfitting, VIMs employs the LoRA \cite{LoRA} adapter. LoRA introduces a small number of trainable parameters into the image encoder, denoising UNet, and image decoder of the LDM model (refer Fig. \ref{fig:model}). This enhances its adaptability while significantly reducing the need for extensive computational resources (refer to the original work \cite{LoRA}). This approach allows VIMs to avoid training from scratch and leverage the pre-trained model, which was trained on a very large vision-language task, and supports single-step inference. To preserve high-resolution details, VIMs incorporates zero-conv skip connection layers \cite{Controlnet} between the image encoder and decoder, facilitating the flow of information across different layers of the network \cite{Controlnet,im2im-Turbo}. These connections improve gradient propagation and maintain the integrity of histopathological structures during the image generation process. 
\vspace{-0.18in}
\subsection{Losses and Adversarial Training}\label{subsec:losstraining}
\vspace{-0.06in}
VIMs utilizes adversarial loss \cite{pix2pix} along with other losses and employs a CLIP-based discriminator \cite{disc,im2im-Turbo}, similar to \cite{SD-Turbo,im2im-Turbo}. The training objective for VIMs involves three key losses: the reconstruction loss $\mathcal{L}_{rec}$, the adversarial (GAN) loss $\mathcal{L}_{adv}$, and the CLIP text-image alignment loss $\mathcal{L}_{clip}$ \cite{clip}. The reconstruction loss $\mathcal{L}_{rec}$, composed of ${L}_2$ and $\mathcal{L}_{lpips}$ losses \cite{lpips}, ensures that the generated IHC images closely match the ground truth. The GAN loss $\mathcal{L}_{adv}$, facilitated by the CLIP-based discriminator, helps produce realistic IHC images by guiding the generator through feedback. The text-image alignment loss $\mathcal{L}_{clip}$ ensures that the output matches the desired IHC marker by aligning the generated images with the provided text prompts. The overall training objective of VIMs is defined as:
\vspace{-0.05in}
\begin{equation}
\mathcal{L}_{total} = \mathcal{L}_{rec} + w_{clip} \mathcal{L}_{clip} + w_{adv} \mathcal{L}_{adv}
\label{eq2}
\end{equation}
\vspace{-0.02in}
where $w_{adv}$ and $w_{clip}$ are the weights for the GAN and CLIP losses, respectively. The objective is to minimize $\mathcal{L}_{total}$ as: $\arg \min_G \mathcal{L}_{total}$.
This adversarial training setup enables VIMs to generate high-fidelity IHC stains from H\&E images while maintaining consistency with the text prompts, allowing for end-to-end training.
\vspace{-0.15in}
\subsection{Inference} \label{subsec:infer}
\vspace{-0.06in}
During inference, the VIMs model generates multiplex IHC images from an input H\&E image and a text prompt. The process involves encoding the H\&E image, integrating it with the encoded text prompt, and decoding the generated latent representation to produce the final IHC image. This allows for the generation of images for multiple IHC markers from the same H\&E image. Our approach addresses the lack of paired H\&E and multiplex IHC data by using conditional text prompts for each marker and incorporating negative samples (images lacking cells positive for CDX2 or CK8/18) in the training dataset. This reduces false positive rates and offers a scalable solution for virtual IHC staining (see Section \ref{sec:exp}).
\vspace{-0.07in}
\section{Experimentation and Discussion} \label{sec:exp}
\vspace{-0.07in}
\subsection{Dataset and Training Details} \label{subsec:datatrain}
\begin{figure}
    \centering
    \includegraphics[clip, trim=1.0cm 7.6cm 1.5cm 0.10cm, scale=0.41]{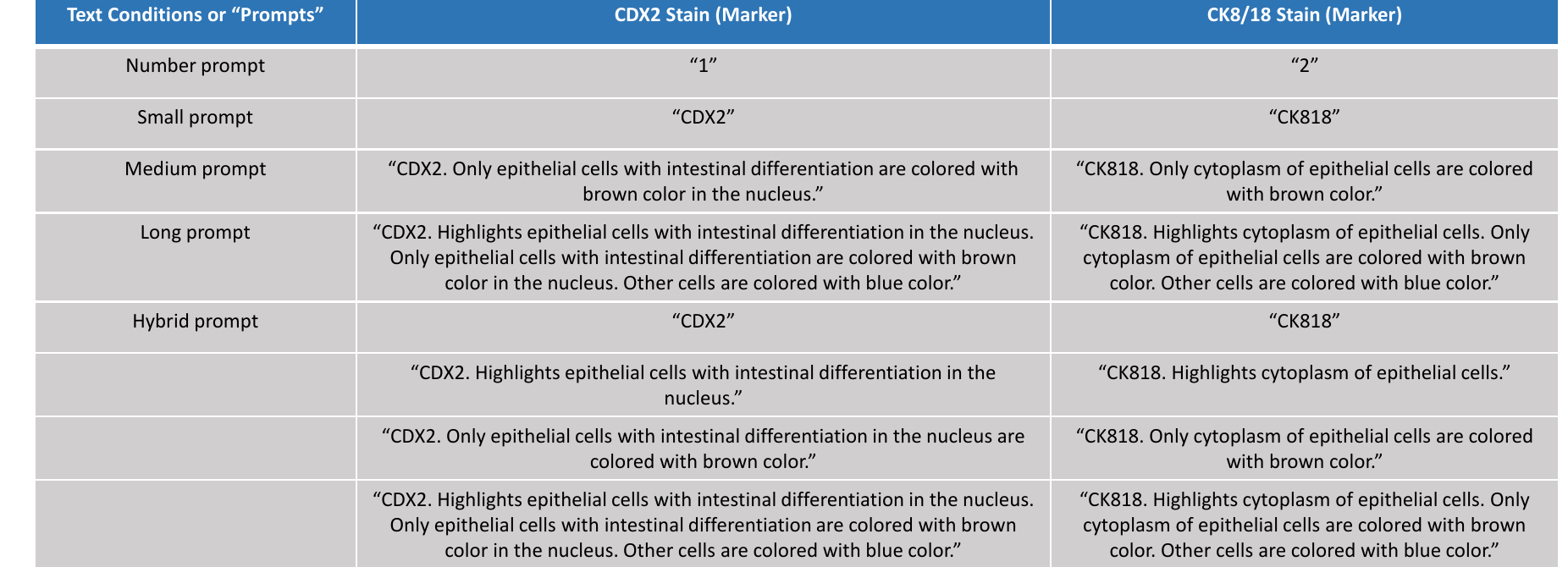}
    \vspace{-0.13in}
    \caption{Overview of various prompts validated by pathologists, used in model training and evaluation to assess their impact on VIMs.}
    \label{fig:prompts}
\vspace{-0.14in}
\end{figure}

\vspace{-0.05in}
This study utilized a uniplex paired dataset, where each H\&E image is pixel-wise registered with a single IHC marker, to train our model. We focused on two gland IHC markers: CDX2, a nuclear marker expressed in the epithelial cells of the colon, and CK8/18, a cytoplasmic marker in epithelial cells. The goal was to generate and evaluate multiplex IHC markers from the same H\&E patch. The model's capability was assessed using a partially paired dataset: paired test data where H\&E is pixel-wise registered to one of the ground truth (GT) IHC markers, and an unpaired test set where the second marker was unavailable. The dataset comprised H\&E WSIs from surveillance colonoscopies of five ulcerative colitis patients. Slides were stained with H\&E, scanned at 0.25 $\mu$m/pixel at 40x magnification, restained by IHC with CDX2 or CK8/18 antibodies, and rescanned. Pixel-level alignment was achieved using ANTSpY \cite{ants}, as detailed in \cite{kataria2023automating}.
For model training, 16,000 patches for CDX2-H\&E and CK8/18-H\&E uniplex pairs were sampled from four patients, and 4,000 patches each for paired testing from a fifth patient. The unpaired test set included H\&E images without the second marker (e.g., CK8/18 stains absent from CDX2-H\&E pairs). Extracted RGB patches were 512x512. Each training sample was paired with prompts, as detailed in Section \ref{subsec:prompt} and Fig.\ref{fig:prompts}. Mixed/hybrid prompts (MxP) were used for VIMs (MxP) model training. Additionally, the dataset for training the UNet \cite{unet} gland segmentation model was created using the pipeline from \cite{kataria2023automating} (refer to GT images Suppl. Fig. \ref{fig:UNET-Seg}). VIMs was trained on an Nvidia A30 GPU with a batch size of 1 for 200,000 steps, with weights $w_{adv} = 0.4$ and $w_{clip} = 4$ as per Eq.\ref{eq2} \cite{im2im-Turbo}. VIMs was effectively fine-tuned with few epochs due to the techniques, including the LoRA adapter, skip connections, and adversarial training. Further details on the evaluation are provided in Section \ref{sec:eval_metric}.
\vspace{-0.1in}
\subsection{Evaluation Methods} \label{sec:eval_metric}
\begin{figure}[t]
    \centering
    \includegraphics[clip, trim=0.05cm 7.7cm 0.1cm 0.10cm, scale=0.29]{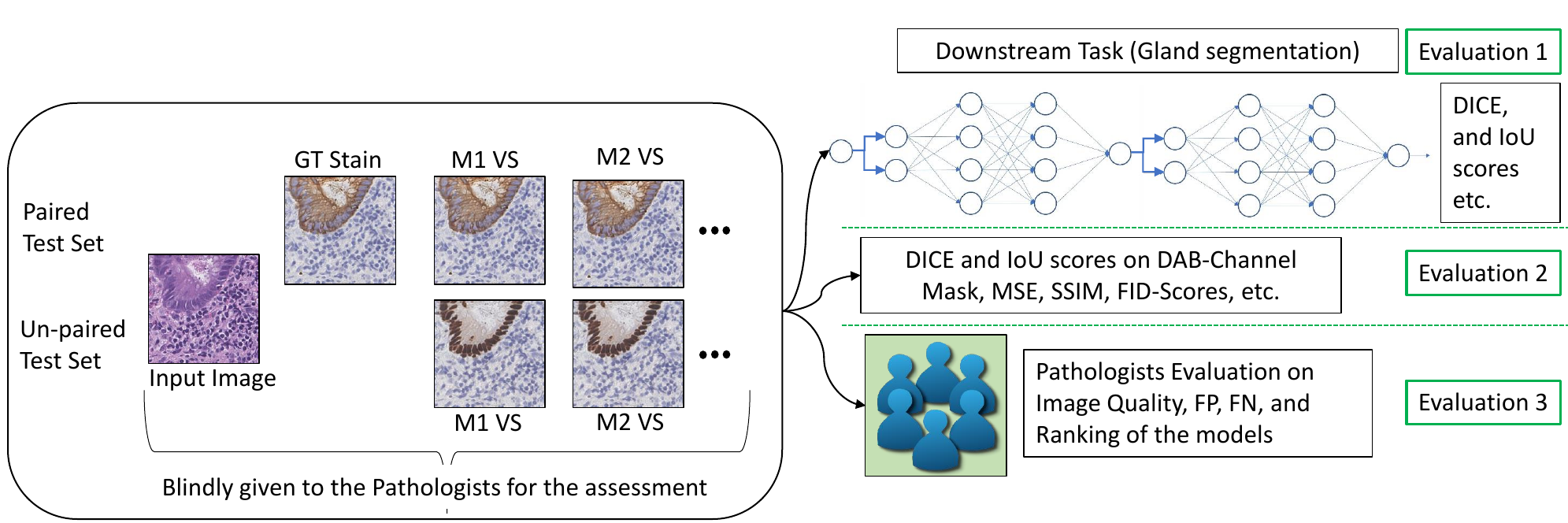}
    \vspace{-0.09in}  
    \caption{Evaluation methodologies for paired and unpaired test sets, including pathologist assessments. For paired test datasets, methods 1, 2, and 3 are employed, while methods 1 and 2 are used for unpaired test datasets. M1 to M4 represent Models 1 to 4, VS: Virtual Stainer.}
    \label{fig:path_eval}
    \vspace{-0.1in}
\end{figure}

\begin{figure}[h]
    \centering
    \begin{subfigure}
        \centering
        \includegraphics[clip, trim=0cm 1.5cm 0cm 1.0cm, scale=0.19]{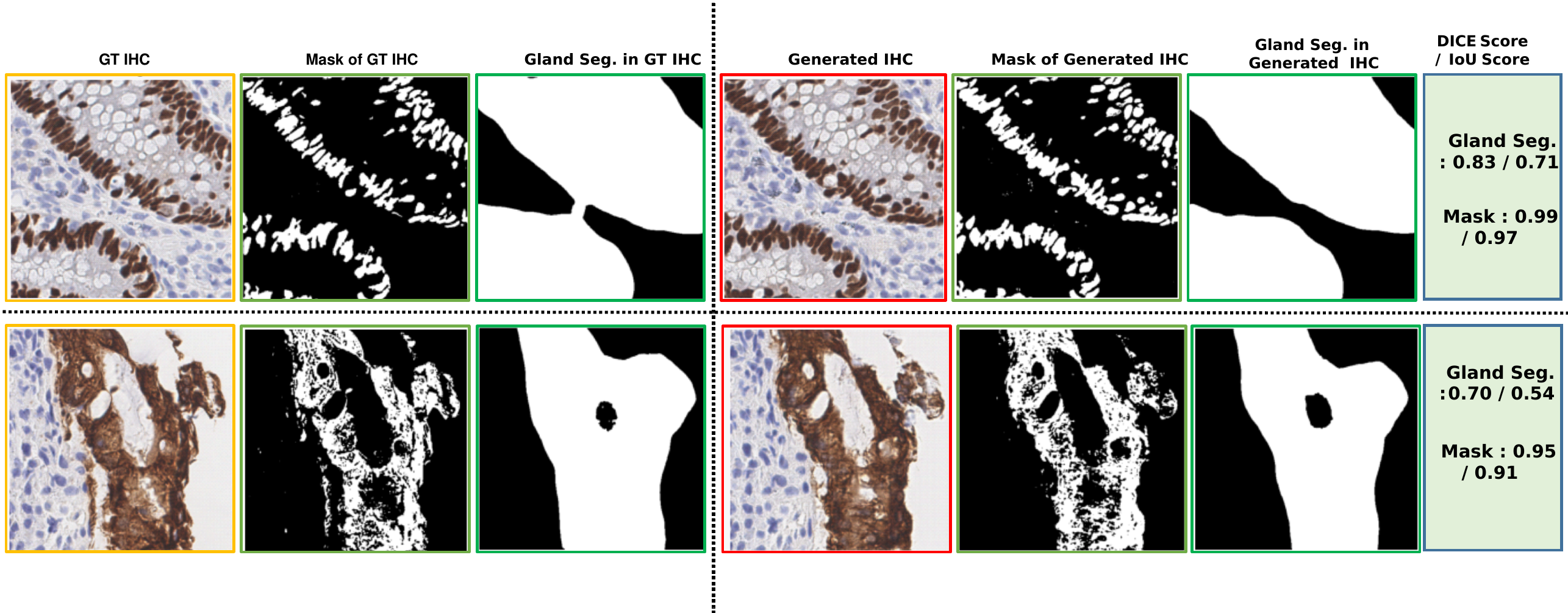}        
        \label{fig:mask_seg_GT}
        \vspace{-0.08in}
    \end{subfigure}    
    \begin{subfigure}
        \centering
        \includegraphics[clip, trim=0.0cm 5.0cm 0.5cm 0.0cm, scale=0.19]{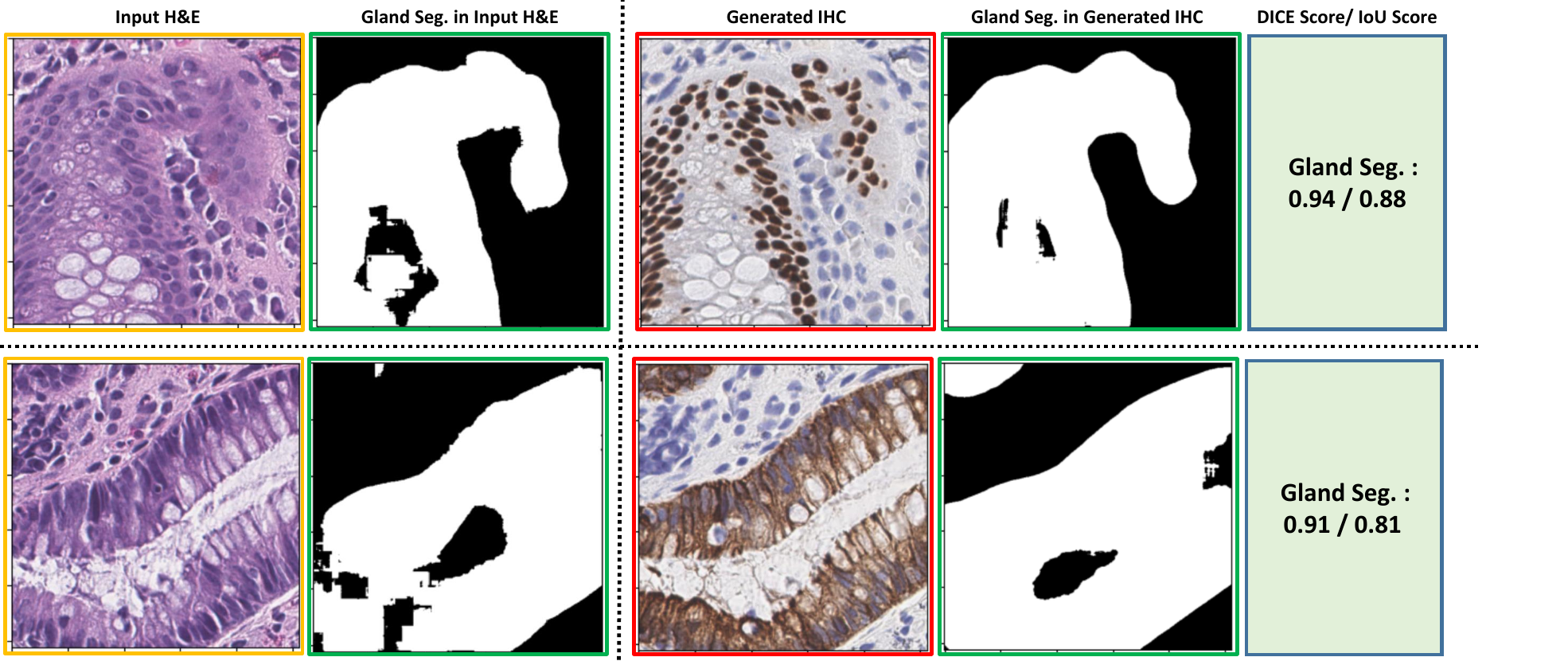}
        \label{fig:mask_seg_noGT}
    \end{subfigure}
     \vspace{-0.05in}
   \caption{Examples of DICE and IoU score evaluations for cases with GT (first 7 columns: row 1 and row 2 for CDX2, CK8/18) and without GT (last 5 columns: two rows for CDX2, CK8/18). Gland segmentation is performed using the trained UNet model \cite{unet}, and mask images are obtained from DAB channel thresholding. Seg.: Segmentation. Outlines: Yellow for GT, Green for DAB Masks and Seg., Red for generated images.}
    \label{fig:mask_seg}
    \vspace{-0.08in}
\end{figure}

\begin{figure}[t]
   \centering
   \includegraphics[clip, trim=3.1cm 1.9cm 0.0cm 1.0cm, scale=0.25]{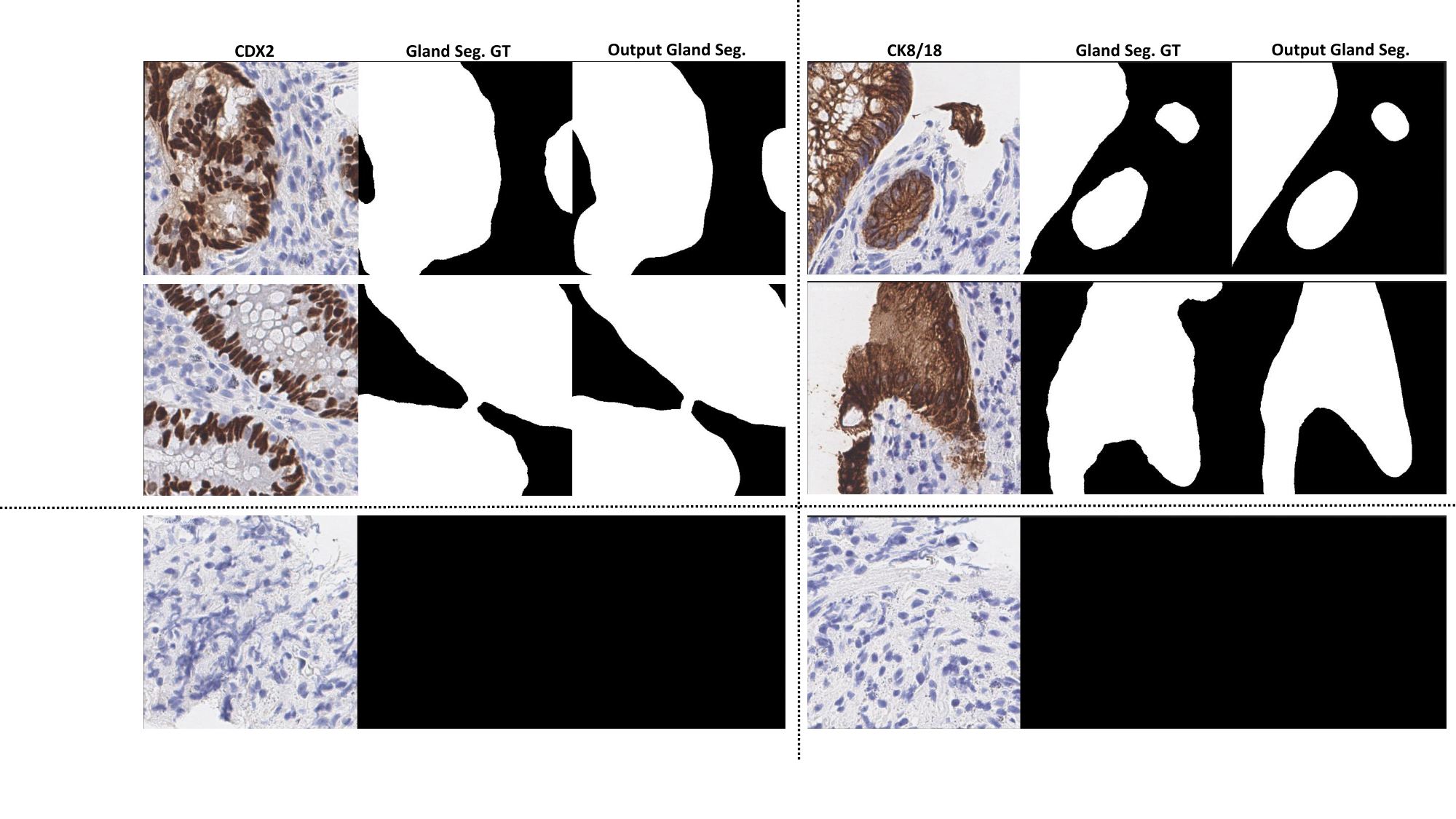}
    \vspace{-0.1in}  
  \caption{Examples of gland segmentation using the trained UNet model on both CDX2 and CK8/18 markers. The third row shows a negative case.}
   \label{fig:UNET-Seg}
   \vspace{-0.3in}
\end{figure}

\vspace{-0.06in}
This study employed three comprehensive quantitative evaluation methods to assess the performance of VIMs, as illustrated in Suppl. Fig. \ref{fig:path_eval}, for both paired and unpaired settings of H\&E and IHC tiles. We aimed to determine if the model (1) correctly reconstructed the glands in the colon and (2) correctly colored the nuclei of epithelial cells when prompted for the CDX2 marker and the cytoplasm of epithelial cells when prompted for the CK8/18 marker. In virtually stained images, unstained glandular epithelial cells were counted as false negatives (FNs) and staining of cells outside of glands as false positives (FPs). Gland segmentation and masking of brown color pixels (DAB-mask) allowed the calculation of  DICE, IoU, and Hausdorff distance (Haus. Dist.) \cite{haus} as the metrics of staining accuracy. The evaluation of VIMs performance included:
\vspace{-0.05in}
\begin{enumerate}
    \item \textbf{Downstream Task (Gland Segmentation) evaluation}: We used a UNet model \cite{unet} trained on gland segmentation tasks (examples in Suppl. Fig.\ref{fig:UNET-Seg}) to evaluate both paired and unpaired test sets. IoU, DICE, and Haus. Dist. \cite{haus} were used to determine the overlap in gland outlines. In the paired setting, the UNet was trained on IHC images with corresponding GT gland segmentation \cite{kataria2023automating}. In the unpaired test set, the UNet was trained on H\&E images with corresponding GT gland segmentation. Examples of this evaluation are shown in Fig.\ref{fig:mask_seg}.
    \item \textbf{Quantitative Metrics}: Standard metrics were used to measure the quality and accuracy of generated IHC images, calculated only for the paired test data. Metrics included DICE, IoU, and Haus. Dist. on DAB-channel masks (Suppl. Fig.\ref{fig:mask_seg}), as well as Mean Squared Error (MSE), Structural Similarity Index (SSIM) \cite{ssim}, and Fréchet Inception Distance (FID) \cite{fid} calculated on the GT and generated IHC images.
     \item \textbf{Qualitative Assessment by Pathologists}: Two board-certified study pathologists visually inspected the generated images to evaluate their quality and accuracy. This involved assessing the overall image quality, identifying FP and FN cells, and ranking the models based on their accuracy. This evaluation was performed for both paired and unpaired test datasets.
\end{enumerate}
\vspace{-0.07in}
In addition to these, qualitative assessments were included for evaluating the model. These evaluation methods collectively provide a robust framework for analyzing and validating the effectiveness of the VIMs model.
\begin{table}[!t]
\centering
\setlength{\tabcolsep}{4pt} % Adjust column separation
\renewcommand{\arraystretch}{1.2} % Adjust row separation
\resizebox{1.02\textwidth}{!}{% Adjust the width here (102% of text width)
\begin{tabular}{l|l|c|c|c|c|c|c|c|c|c}
\hline
\textbf{\begin{tabular}[c]{@{}l@{}}Markers\\ (Paired Test Set)\end{tabular}} & 
        \textbf{Models} & 
        \textbf{MSE($\%$)} $\downarrow$ & 
        \textbf{SSIM($\%$)} $\uparrow$ & 
        \textbf{FID} $\downarrow$ & 
        \multicolumn{3}{c|}{\textbf{Gland Segmentation}} & 
        \multicolumn{3}{c}{\textbf{\begin{tabular}[c]{@{}c@{}}DAB-Channel\\ Mask\end{tabular}}} \\ 
        \hline
         &  &  &  &  & \textbf{DICE($\%$)} $\uparrow$ & \textbf{IoU($\%$)} $\uparrow$ & \textbf{Haus. Dist.} $\downarrow$ & \textbf{DICE($\%$)} $\uparrow$ & \textbf{IoU($\%$)} $\uparrow$ & \textbf{Haus. Dist.} $\downarrow$ \\ 
        \hline
\multirow{4}{*}{CDX2} & Pix2Pix(CDX2) & 12.01 & 63.15 & \textbf{17.86} &  86.66 & 83.64 & 96.80 & 81.19 & 74.63 & 120.54 \\  
 & LDM*(CDX2) & \textbf{11.51} & 67.34 & 18.73 & \textbf{92.81} & \textbf{89.67} & \textbf{63.57} & 84.44 & 76.83 & 106.36 \\  
 & ControlNet & 15.63 & 41.05 & 94.34 & 51.75 & 45.95 & 348.91 & 27.56 & 19.81 & 457.84 \\  
 & \textit{VIMs (SP) (Ours)} & 12.47 & 66.00 & 22.76 & 87.57 & 84.01 & 99.16 & 69.80 & 66.73 & 210.04 \\ \cline{2-11}
 & \multicolumn{1}{|l|}{\textit{\textbf{VIMs (MxP) (Ours)}}} & 12.32 & \textbf{68.11} & 19.21 & 87.83 &  84.79 & 99.66 & \textbf{85.10} & \textbf{76.90} & \textbf{103.45} \\ \hline
\multirow{4}{*}{CK8/18} & Pix2Pix(CK8/18) & 12.69 & 64.75 & 18.25 & 91.96 & 89.55 & 62.35 & 45.46 & 37.72 & 178.87 \\  
 & LDM*(CK8/18) & 12.18 & 67.46 & 17.32 & 91.45 & 88.50 & 131.95 & 36.28 & 27.49 & 216.70 \\  
 & ControlNet & 15.68 & 32.66 & 92.19 & 57.47 & 53.56 & 315.63 & 20.65 & 13.53 & 314.85 \\  
 & \textit{VIMs (SP) (Ours)} & 12.72 & 66.46 & \textbf{17.22} & 89.83 & 88.02 & 90.03 & 28.10 & 19.55 & 275.51 \\ \cline{2-11}
 & \multicolumn{1}{|l|}{\textit{\textbf{VIMs (MxP) (Ours)}}} & \textbf{12.15} & \textbf{68.36} & 18.52 & \textbf{93.43} & \textbf{91.03} & \textbf{60.53} & \textbf{45.67} & \textbf{37.73} & \textbf{171.66} \\ \hline 
\end{tabular}}
\caption{Quantitative analysis of dual IHC stain generation for CDX2 and CK8/18 in the paired test set. Pix2Pix and LDM are uniplex stain generation models, trained separately for CDX2 and CK8/18 markers on a fully paired dataset. VIMs are compared only with models trained on paired data. The ControlNet inference step is 25, while others are 1 step. SP: Small Prompt, MxP: Mixed (Hybrid) Prompt. Bold numbers represent the best scores.}
\label{tab:comparison_paired_models}
\vspace{-0.3in}
\end{table}
\vspace{-0.1in}
\subsection{Results} \label{subsec:results}
\vspace{-0.06in}
The proposed model VIMs (MxP) was evaluated on paired and unpaired test data for markers CDX2 and CK8/18, using the methods outlined in Section \ref{sec:eval_metric}. It was compared with Pix2Pix \cite{pix2pix}, LDM* \cite{latentsapcemodel}, ControlNet \cite{Controlnet}, and VIMs (SP) (using less informative prompts, Small Prompts). LDM* shares a similar structure to VIMs, including LoRA adapters and adversarial loss, but without prompts. ControlNet used image and prompt conditioning separately and was trained from scratch with 50 steps and inference with 25 steps, lacking image-prompt direct association, pretrained LVDM, and adversarial training. Unlike VIMs, which is a multiplex staining model, Pix2Pix and LDM* are trained separately for each marker.

\textbf{Quantitative Analysis}: \textbf{(i) Downstream Task Analysis}: CDX2 and CK8/18 were evaluated for gland segmentation. Table \ref{tab:comparison_paired_models} (paired test set) and Table \ref{tab:comparison_unpaired} (unpaired test set, Suppl.) show VIMs (MxP) outperforming other methods in CK8/18 segmentation. LDM showed better CDX2 segmentation, suggesting benefits in separate marker training. ControlNet struggled due to lack of pretrained LVDM, necessitating further optimization and data. Pix2Pix gives comparable results in all cases. \textbf{(ii) Quantitative Metrics Analysis}: Metrics (MSE, SSIM, FID scores, and DAB-Mask scores) were evaluated on paired data (Table \ref{tab:comparison_paired_models}). VIMs (MxP) exhibited superior SSIM scores for both markers, indicating better structural fidelity than Pix2Pix, LDM, and ControlNet. For MSE and FID scores, VIMs (MxP) showed comparable results to Pix2Pix and LDM models, demonstrating its capability in generating high-quality images similar to specialized models trained separately for each marker. VIMs also outperformed in DAB-Mask scores, despite challenges with the less prominent brown color in CK8/18.
% \vspace{-0.10in}
 \begin{figure}[!t]
    \centering
    \includegraphics[clip, trim=0.75cm 0.00cm 0.00cm 0.00cm, scale=0.44]{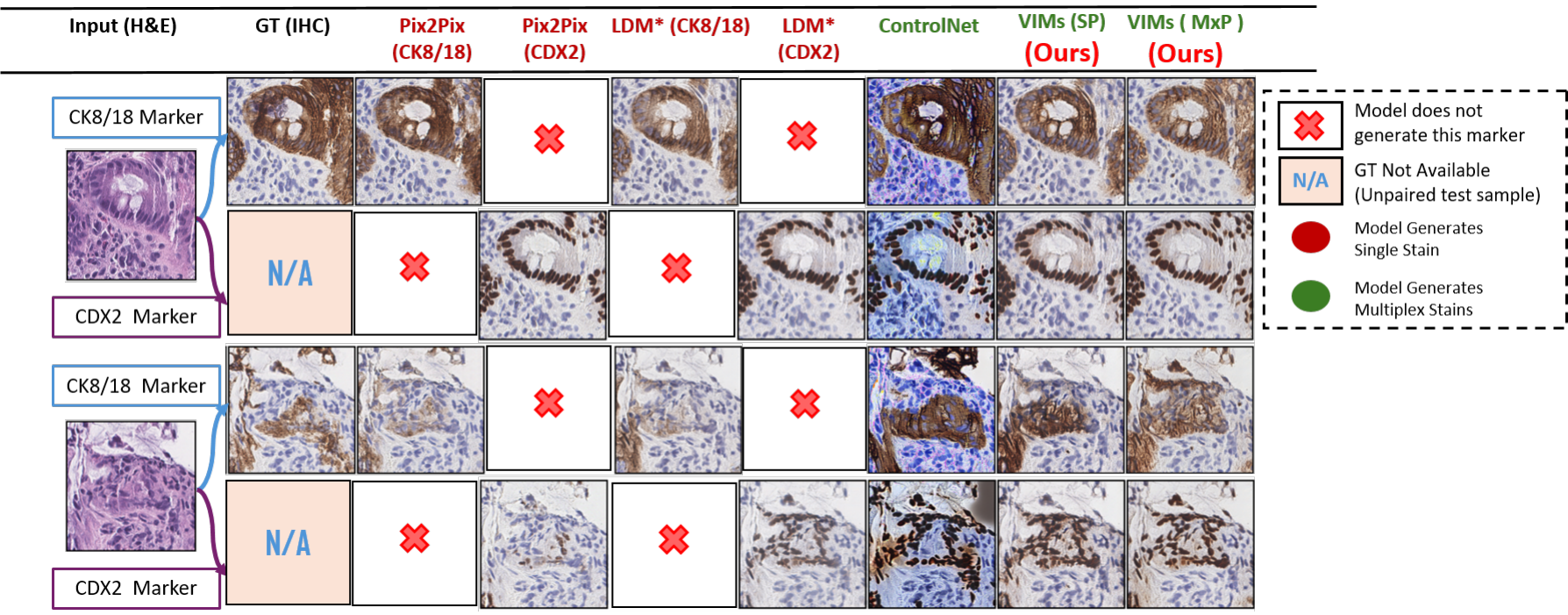}
    \vspace{-0.07in}     \caption{\textbf{Visualization of Multiplex IHC stain generation} on the test set with the CK8/18 GT marker. VIMs generates visually realistic images, accurately highlighting both markers, and performs well across various cases, including difficult samples like the 2nd H\&E input example.}
    \label{fig:qual:ck818}
    \vspace{-0.22in}
\end{figure}
\textbf{(iii) Qualitative and Quantitative Assessment by Pathologists}: This evaluation is particularly significant for cases absent of GT for the generated markers. Two pathologists evaluated 50 challenging H\&E images with generated CDX2 and CK8/18 (refer Table \ref{tab:path_eval}) for both unpaired and paired test sets. Models M1, M2, M3, and M4, are Pix2Pix \cite{pix2pix}, LDM \cite{latentsapcemodel}, VIMs (SP), and VIMs (MxP), respectively. Pix2Pix and LDM are trained separately on CDX2 and CK8/18 markers. M4 achieved top rankings for image fidelity and marker localization accuracy, showing fewer FN compared to other models. M1 performed well in minimizing FP for CDX2, while M4 also showed competitive results. Overall, pathologists rated VIMs (M4) with the highest rank for both paired (only CDX2) and unpaired scenarios.

\begin{figure}[!htb]
    \centering
     \includegraphics[clip, trim=0.0cm 5.0cm 0.5cm 0.00cm, scale=0.44]{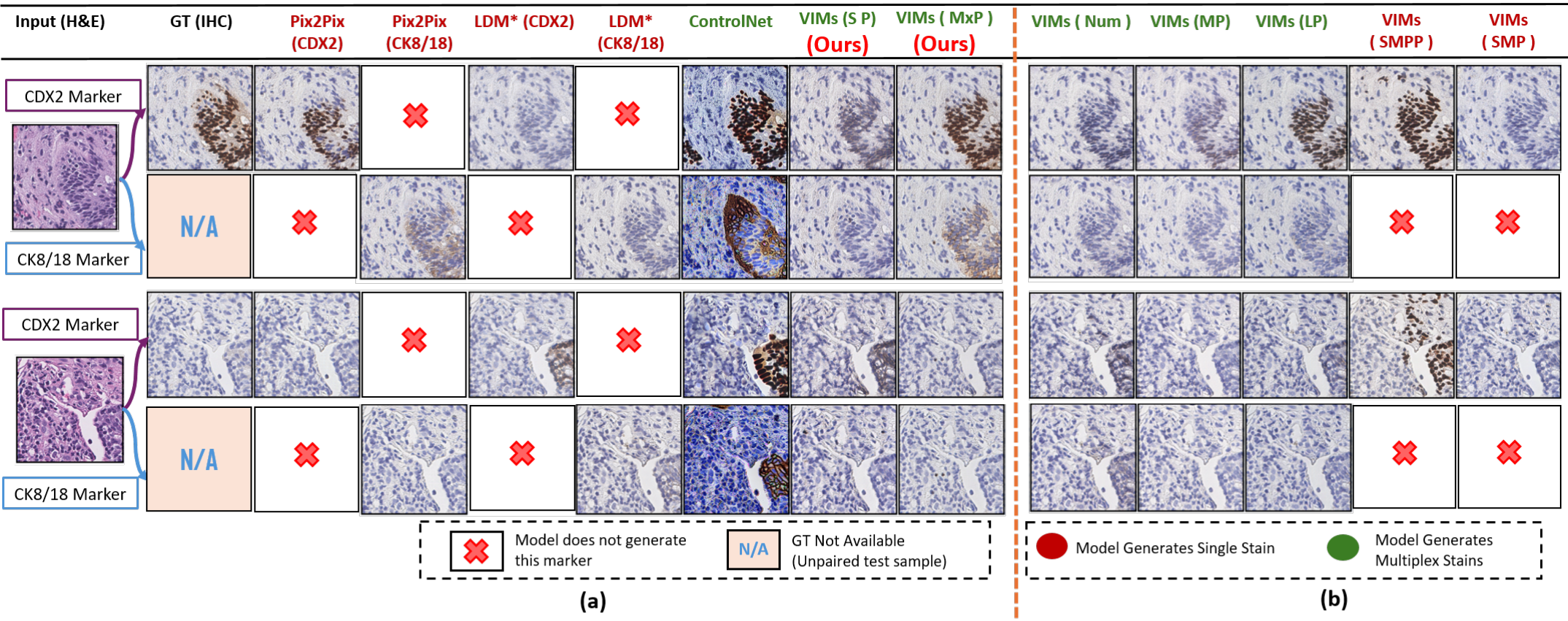}
    \vspace{-0.16in}
 \caption{\textbf{Qualitative results for Multiplex IHC stain generation} on the test dataset with GT for the CDX2 marker.(a) Comparison of VIMs-generated IHC images with Pix2Pix \cite{pix2pix}, LDM \cite{latentsapcemodel}, and ControlNet \cite{Controlnet}. ControlNet inference step is 25, while others are 1 step. The proposed VIMs model generates visually realistic and accurately highlighted images for both CDX2 and CK8/18 markers, performing well across various case types, including negative samples such as the 2nd H\&E input example.(b) Impact of conditioned prompt types and sample types on VIMs performance.}
    \label{fig:qual:cdx2}
    \vspace{-1.8in}
\end{figure}

\textbf{Qualitative Analysis}: Qualitative analysis (Fig.\ref{fig:qual:ck818}, Suppl. Fig.\ref{fig:qual:cdx2}) supported pathologists' assessments. VIMs (MxP) accurately highlighted CDX2 and CK8/18 markers, even with small prompts (VIMs SP). In negative H\&E samples, where glandular structures were absent, VIMs effectively generated negative IHCs (Fig.\ref{fig:qual:cdx2}), whereas other models struggled. Similar performance for positive samples is shown in the second example in Fig.\ref{fig:qual:ck818}. This robust performance underscores VIMs' efficacy in generating multiplex IHC images from a single trained model.

Overall, these evaluations demonstrate VIMs (MxP)'s capability in generating multiple markers with high fidelity, closely matching GT feature spaces, exhibiting fewer pixel-wise variations, and accurately positioning structures in H\&E samples.

\begin{table}[!t]
\centering
\setlength{\tabcolsep}{5pt} % Adjust column separation
\renewcommand{\arraystretch}{1.1} % Adjust row separation
\resizebox{0.7\textwidth}{!}{% Adjust the width here (70% of text width)
\begin{tabular}{l|l|cccc|cccc|cccc}
\hline
\textbf{Markers} & \textbf{Pathologists} & \multicolumn{4}{c|}{\textbf{False Positive (FP)(\%)} $\downarrow$} & \multicolumn{4}{c|}{\textbf{False Negative (FN)(\%)} $\downarrow$} & \multicolumn{4}{c}{\textbf{Model Ranking} $\downarrow$} \\ 
\hline
 &  & M1 & M2 & M3 & \textbf{M4} & M1 & M2 & M3 & \textbf{M4} & M1 & M2 & M3 & \textbf{M4} \\ 
\hline
\multirow{3}{*}{\shortstack[l]{Unpaired\\ CDX2 Test}} & P1 & 10.00 & 11.00 & 11.00 & 10.80 & 12.00 & 10.20 & 10.00 & 10.40 & 3 & 2 & 4 & 1 \\ 
 & P2 & 10.20 & 10.60 & 11.40 & 10.80 & 13.10 & 10.20 & 10.00 & 10.00 & 2 & 3 & 4 & 1 \\ 
\cline{2-14}
 & Ave. Score & \textbf{10.10} & 10.80 & 11.20 & 10.80 & 12.55 & 10.20 & \textbf{10.00} & 10.20 & 2 & 2 & 3 & \textbf{1} \\ 
\hline
\multirow{3}{*}{\shortstack[l]{Unpaired\\ CK8/18 Test}} & 
P1 & 11.40 & 11.80 & 11.60 & 10.80 & 16.40 & 12.00 & 12.60 & 10.80 & 3 & 4 & 2 & 1  \\
& P2 & 10.00 & 10.20 & 10.80 & 10.20 & 16.60 & 13.40 & 11.80 & 11.20 & 4 & 3 & 2 & 1 \\ 
\cline{2-14}
 & Ave. Score & 10.70 & 11.00 & 11.20 & \textbf{10.50} & 16.50 & 12.70 & 12.20 & \textbf{11.00} & 3 & 3 & 2 & \textbf{1} \\ 
\hline
\multirow{3}{*}{\shortstack[l]{Paired\\ CDX2 Test}} & P1 & 10.00 & 10.00 & 10.20 & 10.20 & 17.70 & 16.00 & 14.10 & 11.80 & 3 & 2 & 4 & 1 \\ 
 & P2 & 10.20 & 10.40 & 11.40 & 10.20 & 21.10 & 17.20 & 14.30 & 11.00 & 2 & 3 & 4 & 1 \\ 
\cline{2-14}
 & Ave. Score & \textbf{10.10} & 10.20 & 10.80 & 10.20 & 19.40 & 16.60 & 14.20 & \textbf{11.40} & 2 & 2 & 3 & \textbf{1}  \\ 
\hline
\end{tabular}
}
\caption{Pathologists' evaluation on the unpaired and paired test sets with a single inference step. Model rankings (Range: \{1, 2, 3, 4\}). P1 and P2 are Pathologists 1 and 2, respectively; Avg.: Average; M1: Pix2Pix \cite{pix2pix}; M2: LDM \cite{latentsapcemodel}; M3: VIMs (SP) (Ours); \textbf{M4}: \textbf{\textit{VIMs (MxP) (Ours)}}. Pix2Pix and LDM are trained separately on CDX2 and CK8/18 markers.}
\label{tab:path_eval}
\vspace{-0.30in}
\end{table}
\vspace{-0.1in}
\subsection{Ablation Study} \label{sec:path_eval} \label{subsec:ablation}
\vspace{-0.07in}
In addition to the evaluations in Section \ref{subsec:results}, this study conducted an ablation analysis on various prompts used in the VIMs models for both markers. We examined VIMs on different data types, such as training without negative samples on a single marker (CDX2), Single-Marker Positive Prompts (SMPP), and combined positive and negative samples on a single marker (SMP). We also assessed the impact of different prompt conditions, including number prompts (Num) and varying lengths (small (SP), medium (MP), long (LP), mixed/hybrid (MxP)). Results from Suppl. Table \ref{tab:comparison_paired_prompt} and Fig.\ref{fig:qual:cdx2} demonstrate that VIMs (MxP) outperforms other variations, including VIMs (SP)(Tables \ref{tab:comparison_paired_models} and \ref{tab:comparison_unpaired}). SMPP generated more FP than SMP, highlighting the importance of including negative samples for better accuracy and visual quality. The comparison of VIMs (MxP) to VIMs (Num) shows the importance of informative prompt conditioning. VIMs (MP) performed similarly to VIMs (MxP), but the latter benefits from requiring less informative prompts during inference. VIMs (LP) needs further optimization of the prompt encoder. Overall, VIMs (MxP) consistently delivered high-quality results for both markers, leveraging hybrid prompts for robust performance, highlighting its efficiency in generating multiplex IHC images from a single trained model.
\vspace{-0.14in}
\section{Conclusion and Future Work} \label{sec:conclusion}
\vspace{-0.09in}
This study introduced Virtual Immunohistochemistry Multiplex staining (VIMs), using a text-conditioned single-step DM to generate multiple IHC stains from a single H\&E sample. VIMs, the first to adapt a LVDM for virtual IHC multiplexing with adversarial learning. Our approach addresses the challenge of unavailable paired H\&E and multiplex IHC data, providing a scalable solution. Extensive experiments showed VIMs outperforming traditional GAN-based models, achieving high fidelity in IHC stain generation. Pathologists' evaluation confirmed its ability to produce diagnostically relevant IHC images. Efficient training with the LoRA and incorporating negative samples to reduce false positives further enhanced it's performance. Future work includes expanding VIMs to more challenging IHC markers like CD3 and CD20, improving text conditioning with advanced NLP techniques, and integrating VIMs into other clinical uses. Investigating robustness and generalizability across diverse datasets and developing automated evaluation metrics aligned with pathologist assessments are crucial next steps. Enhancing model efficiency for real-time applications and eliminating the need for uniplex paired data will also be explored. VIMs advances virtual IHC staining, offering a scalable method for generating multiplex IHC stains from a single H\&E sample.
\vspace{-0.2in}
\section*{Acknowledgements}
\vspace{-0.13in}
We thank the Department of Pathology and the Kahlert School of Computing at the University of Utah for their support of this project.  This work was supported in part by the U.S. National Science Foundation (NSF) through award 2217154.
\vspace{-0.08in}
% 
% ---- Bibliography ----
%
\vspace{-0.10in}
 \bibliographystyle{splncs04}
 % \vspace{-0.08in}
 \bibliography{cam-ready.bib}
 \newpage
 \section{Supplementary}
 \vspace{-0.2in}
\begin{table}[h]
\centering
\setlength{\tabcolsep}{4pt} % Adjust column separation
\renewcommand{\arraystretch}{1.2} % Adjust row separation
\resizebox{1.02\textwidth}{!}{% Adjust the width here (102% of text width)
\begin{tabular}{l|l|c|c|c|c|c|c|c|c|c}
\hline
\textbf{\begin{tabular}[c]{@{}l@{}}Markers\\ (Paired Test Set)\end{tabular}} & 
        \textbf{Models} & 
        \textbf{MSE($\%$)} $\downarrow$ & 
        \textbf{SSIM($\%$)} $\uparrow$ & 
        \textbf{FID} $\downarrow$ & 
        \multicolumn{3}{c|}{\textbf{Gland Segmentation}} & 
        \multicolumn{3}{c}{\textbf{\begin{tabular}[c]{@{}c@{}}DAB-Channel\\ Mask\end{tabular}}} \\ 
\hline
         &  &  &  &  & \textbf{DICE($\%$)} $\uparrow$ & \textbf{IoU($\%$)} $\uparrow$ & \textbf{Haus. Dist.} $\downarrow$ & \textbf{DICE($\%$)} $\uparrow$ & \textbf{IoU($\%$)} $\uparrow$ & \textbf{Haus. Dist.} $\downarrow$ \\ 
\hline
\multirow{4}{*}{CDX2} & VIMs (Num) & 12.35 & 65.92 & 19.91 & 81.70 & 77.84 & 102.28 & 69.69 & 62.57 & 192.96 \\   
 & VIMs (MP) & \textbf{11.58} & 68.18 & 19.55 & 86.90 & 83.86 & 99.37 & 78.02 & 71.39 & 148.83 \\ 
 & VIMs (LP) & 13.04 & \textbf{76.09} & 19.58 & 80.64 & 75.11 & 156.33 & 70.51 & 63.60 & 139.39 \\ 
 & VIMs (SMPP) & 12.44 & 58.23 & 68.05 & 44.08 & 40.29 & 424.16 & 39.19 & 32.31 & 423.02 \\ 
 & VIMs (SMP) & 11.91 & 66.38 & 21.26 & 80.41 & 74.41 & 179.81 & 65.33 & 58.49 & 227.45 \\ \cline{2-11}
 & \multicolumn{1}{|l|}{\textit{\textbf{VIMs (MxP) (Ours)}}} & 12.32 & 68.11 & \textbf{19.21} & \textbf{87.83} &  \textbf{84.79} & \textbf{99.66} & \textbf{85.10} & \textbf{76.90} & \textbf{103.45}\\ \hline
\multirow{4}{*}{CK8/18} & VIMs (Num) & 12.22 & 66.60 & 18.93 & 92.20 & 88.79 & 74.80 & 41.80 & 34.38 & 185.61 \\  
 & VIMs (MP) & \textbf{11.73} & 68.25 & 18.58 & 93.48 & 90.90 & 67.58 & 40.70 & 34.86 & 193.88 \\  
 & VIMs (LP) & 13.38 & 66.97 & \textbf{16.55} & 93.03 & 90.11 & 66.38 & 32.13 & 23.46 & 249.60 \\ 
 \cline{2-11}
 & \multicolumn{1}{|l|}{\textit{\textbf{VIMs (MxP) (Ours)}}} & 12.15 & \textbf{68.36} & 18.52 & \textbf{93.43} & \textbf{91.03} & \textbf{60.53} & \textbf{45.67} & \textbf{37.73} & \textbf{171.66} \\ \hline
\end{tabular}
}
\caption{Analysis of the impact of conditioned prompt types and sample types on VIMs performance for both markers in the paired test set with 1 inference step. MxP: Mixed (Hybrid) Prompt, Num: Number Prompt, MP: Medium Prompt, LP: Long Prompt, SMPP: Single-Marker Positive Prompt, SMP: Single-Marker Prompt. Bold numbers represent the best scores.}
\label{tab:comparison_paired_prompt}
\end{table}
\vspace{-0.6in}
\begin{table}[h]
\centering
\setlength{\tabcolsep}{4pt} % Adjust column separation
\renewcommand{\arraystretch}{1.2} % Adjust row separation
\resizebox{0.6\textwidth}{!}{% Adjust the width here (60% of text width)
\begin{tabular}{l|l|c|c|c}
\hline
\textbf{\begin{tabular}[c]{@{}l@{}}Markers\\ (Unpaired Test Set)\end{tabular}} & \textbf{Models} & \textbf{DICE($\%$)} $\uparrow$ & \textbf{IoU($\%$)} $\uparrow$ & \textbf{Haus. Dist.} $\downarrow$ \\ 
\hline
\multirow{5}{*}{CDX2} & Pix2Pix(CDX2) & 85.83 & 80.79 & 90.99 \\  
 & LDM*(CDX2) & 84.88 & 79.78 & 95.79 \\  
 & ControlNet & 45.59 & 38.61 & 355.15 \\  
 & \textit{VIMs (SP) (Ours)} & 81.31 & 76.29 & 122.16 \\  \cline{2-5}
 & \textit{\textbf{VIMs (MxP) (Ours)}} & \textbf{86.30} & \textbf{81.38} & \textbf{85.12} \\ 
\hline
\multirow{5}{*}{CK8/18} & Pix2Pix(CK8/18) & 86.41 & \textbf{83.96} & 94.53 \\  
 & LDM*(CK8/18) & 85.17 & 81.40 & 131.95 \\  
 & ControlNet & 52.69 & 47.60 & 327.80 \\  
 & \textit{VIMs (SP) (Ours)} & 83.51 & 79.89 & 111.94 \\  \cline{2-5}
 & \textit{\textbf{VIMs (MxP) (Ours)}} & \textbf{86.62} & 83.05 & \textbf{94.44} \\ 
\hline
\end{tabular}
}
\caption{Analysis of Multiplex IHC stain generation for both markers in the unpaired test set. Models marked with * are trained similarly to VIMs for fair comparison. Pix2Pix \cite{pix2pix} and Latent Diffusion Model (LDM) \cite{latentsapcemodel} are uniplex stain generation models. ControlNet \cite{Controlnet} inference step is 25, while others are 1 step. DICE and IoU scores are calculated as illustrated in Fig.\ref{fig:mask_seg}. SP: Small Prompt. Bold numbers indicate the best scores.}
\label{tab:comparison_unpaired}
\vspace{-0.6in}
\end{table}
\vspace{-0.17in}
\end{document}